# Molecular signals from primordial clouds at high redshift

*Roberto Maoli*

DEMIRM, Observatoire de Paris-Meudon

61, avenue de l'Observatoire, 75014 Paris, France

and

Laboratoire de Radioastronomie Millimétrique, Ecole Normale Supérieure

24 rue Lhomond, 75005 Paris, France

*Valerio Ferrucci*

Dipartimento di Fisica, Università di Roma "La Sapienza"

P.le Aldo Moro 2, 00185 Roma, Italy

*Francesco Melchiorri*

Dipartimento di Fisica, Università di Roma "La Sapienza"

P.le Aldo Moro 2, 00185 Roma, Italy

*Monique Signore*

DEMIRM, Observatoire de Paris-Meudon

61, avenue de l'Observatoire, 75014 Paris, France

and

Laboratoire de Radioastronomie Millimétrique, Ecole Normale Supérieure

24 rue Lhomond, 75005 Paris, France

*Danilo Tosti*

Dipartimento di Fisica, Università di Roma "La Sapienza"

P.le Aldo Moro 2, 00185 Roma, Italy

# ABSTRACT

The possibility to detect cosmological signals from the post–recombination Universe is one of the main aims of modern cosmology. In a previous paper we emphasized the role that elastic resonant scattering through LiH molecules can have in dumping primary CBR anisotropies and raising secondary signals. Here we extend our analysis to all the evolutionary stages of a primordial cloud, starting with the linear phase, through the turn–around and to the non linear collapse. We have done calculations for proto–clouds in a CDM scenario and, more generally, for a set of clouds with various masses and various turn–around redshifts, in this case without referring to any particular structure formation scenario. We found that the first phase of collapse, for $t/t_{free-fall} = 0.05 \div 0.2$ is the best one for simultaneous detection of the first two LiH rotational lines. The observational frequency falls between 30 and 250 $GHz$ and the line width $\frac{\Delta\nu}{\nu}$ is between $10^{-5}$ and $10^{-4}$. As far as we know this is the most favourable process to detect primordial clouds before they start star formation processes.

*Subject Headings:* cosmic microwave background – early universe – galaxies: formation – molecular processes.

# 1–Introduction

If we don't suppose a reionization epoch, the Universe became transparent for the cosmic radiation starting from $z \sim 1000$, when the residual ionization wasn't enough to ensure the coupling between the photons and the baryonic matter through the Thomson scattering. As a consequence the only information reaching us comes from a very primordial epoch when the Universe was almost perfectly homogenous; moreover the finite depth of the last scattering surface statistically erases perturbations of angular dimension less than 8 $\Omega^{1/2} h^{-1}$ *arcmin* so that the perturbations at the scales relative to the present structures are not observable.

The existence of this "dark age" between redshifts 1000 and 5 is further annoying since it does not exist up to today any theory about cosmic structures formation that fits the available experimental data coming from the Cosmic Background Radiation (CBR) anisotropies experiments, the maps and statistical description of the large scale distribution of galaxies, and the measurements of the peculiar motions of galaxies. In particular, the standard cold dark matter (CDM) model normalized with COBE–DMR results predicts too much power on small scales and too large pairwise galaxy velocities.

We can understand the importance of getting informations from these intermediate redshifts, where the protostructures have left their linear evolution phase to go into a non linear collapse. The forming proto–structures can leave their imprint on the CBR through different processes: Thomson scattering by ionized matter having a peculiar velocity (Ostriker & Vishniac 1986; Vishniac 1987; Hu, Scott & Silk 1993), inverse Compton scattering by heated gas (Cole & Kaiser 1988; Schaeffer & Silk 1988), thermal emission of heated dust at high redshift (Bond, Carr & Hogan 1991) and gravitational effect of growing non–linear structures (Rees & Sciama 1968; Martìnez–Gonzàlez,

Sanz & Silk 1990) . The first three processes refer to an advanced phase of the collapse or need a precedent generation of heating sources, while the last one produces CBR fluctuation only if very large structures are involved.

A different source of secondary anisotropies can be the elastic resonant scattering by primordial molecules.

This component of the cosmic medium was first considered as the main cooling agent in the dust–free clouds responsible for primordial star formation (Silk 1983; Palla, Salpeter & Stahler 1983; Lepp & Shull 1984).

The cosmological consequences of the presence of molecules at high redshifts were suggested by Dubrovich (1977). The elastic resonant scattering can damp CBR primary anisotropy. If the scattering source has a component of peculiar motion along the line of sight a rotational and roto–vibrational like spectrum is produced. Beyond its spectral nature, the peculiarity of this kind of secondary anisotropy is the no needing of the presence of hot gas or ionized matter to be produced. For this reason the molecular signal is very promising for investigating primordial structures during their linear phase or the first phase of the collapse.

In a previous paper (Maoli, Melchiorri & Tosti 1994, hereafter MMT) we developed a complete computation of the optical depth relative to the different primordial molecules. LiH is the best candidate to produce cosmological effects.

Recently Dubrovich (1993) proposed $H_2D^+$ as a possible candidate for the production of secondary anisotropies. He supposed a total conversion of deuterium in $H_2D^+$ for $z \leq 50$; this is possible only with a high concentration of free electrons ($n_e \approx 10^{-3}$) that implies the existence of an unidentified high redshift heating source.

Palla, Galli & Silk (1995) extended our optical depth computation to the case of a deuterium rich universe ($(D/H)_p = 1.9 \cdot 10^{-4}$) confirming the importance of LiH, but finding reduced effects for the primary anisotropies attenuation. They also found that the $H_2D^+$ contribution to the total optical depth is negligible ($\tau < 10^{-9}$).

In MMT we also calculated the secondary signal associated to a moving proto–cloud during its linear evolution that is when it follows the expansion of the Universe, in this paper we will extend our analysis to the turn–around and the non–linear collapse phase.

In section 2 we describe the assumptions for the calculation in the non–linear collapse, following a thermodynamical approach; in section 3 we describe the expected signal from such objects; in subsection 4.1 we analyze the obtained results in the framework of the standard CDM model; finally in subsection 4.2 we generalize our analysis to proto–clouds of different masses and turn–around redshifts, without referring to any structure formation model.

## 2–The phases of perturbation evolution

The problem of galaxy formation in the gravitational instability hypothesis has been studied by many authors (see, for instance, Peebles 1993). It is possible to schematize three stages of the evolution of perturbations:

1) linear evolution: the initial perturbation $\frac{\delta\rho}{\rho}$ grows algebraically ($\propto \frac{1}{1+z}$) with time, following the expansion of the universe;

2) turn–around: the perturbation has grown up to about $\frac{\delta\rho}{\rho} = \left(\frac{3\pi}{4}\right)^2$. The radius of the fluctuation (for simplicity assumed of spherical shape) has reached its maximum value; the collapse compensates the expansion of the universe and the perturbation appears non–moving;

3) non–linear collapse: it is the collapsing phase properly said; density, temperature and pressure increase very quickly up to the beginning of thermal instability and the evolution of the perturbation towards the bound structures we observe today.

We summarize our assumptions for the non–linear collapse. We adopt the standard cosmological model with zero cosmological constant, we set $\Omega = 1$, $\omega_b = 0.1$, $h_{100} = 0.75$; we use the thermodynamical approach; as a toy model we consider the collapse of a non–rotating spherical perturbation of primordial chemical composition. We ignore the effect of the cloud opacity as well as the magnetic field.

The derivation of the evolution laws of the cloud collapse starts from the first law of thermodynamics

$$dU = -PdV + dQ \qquad (1)$$

where $U = 3/2 N k_B T$ is the internal energy, $P = N k_B T/V$ is the pressure, $dQ = -\Lambda V dt$ and $V = N X \mu (r/r_{ta})^3 / n_{H,ta}$; $\Lambda$ is the cooling function, i.e. the energy loss per unit volume and unit time. We only consider the contribution of $H_2$ molecule (Lepp and Shull 1983), we set the initial abundance of molecular hydrogen $h_2 = 1.2 \times 10^{-6}$, following Lepp & Shull (1984) and the initial ionization fraction $x_e = 3 \times 10^{-5} \Omega_0^{1/2}/(\omega_b h)$ (Padmanabhan 1993). It is possible to take into account other effects in the expression of the cooling function as, for example, the contributions of other molecules (HD and LiH) or the Compton coupling between the free electrons and the radiation, nevertheless in our case we found that these terms are negligible.

Following Lahav (1986), we introduce the adimensional variables $f = T/T_{ad}$ and $x = r/r_{ta}$ (where $T_{ad} = T_{ta}(r/r_{ta})^{-2}$ is the cloud temperature for an adiabatic collapse) and we get:

$$\frac{df}{dt} = -\frac{2}{3} \frac{X\mu}{k_B n_{H,ta} T_{ta}} \Lambda x^5 \qquad (2).$$

where $X$ is the primordial hydrogen mass fraction, $\mu$ is the mean molecular weight and $ta$ means that quantities are relative to the turn–around time.

The evolution of the cloud radius is described by the equation

$$\frac{d^2 r}{dt^2} = \frac{5 k_B T}{\mu m_H r} - \frac{GM}{r^2} \qquad (3)$$

given by the conservation of energy.

In order to evaluate the cooling factor we need to follow the behaviour of the chemical abundances involved in the cooling processes. Using the reactions reported in table 1 and applying a steady–state approximation to the concentration of $H^-$, we obtained the following equations:

$$\frac{dn_H(t)}{dt} = -3n\frac{\dot{r}}{r}$$
$$\frac{dx_e}{dt} = -k_3 n_H x_e^2 \qquad (4)$$
$$\frac{dh_2}{dt} = \frac{k_1 k_2 n_H^2 x_e}{k_2 n_H + k_4}$$

where $x_e = n_e/n_H$, and $h_2 = n_{H_2}/n_H$.

The total description of the collapse is obtained by solving the differential equation system (2) to (4) with the initial conditions given at the turn–around time. Following the classical approach (Gunn & Gott 1972) we take:

$$\rho_{ta} = \left(\frac{3\pi}{4}\right)^2 \Omega_0 \omega_B \rho_{cr}(1+z_{ta})^3$$
$$1 + z_{ta} = \left(\frac{3\pi}{4}\right)^{-\frac{2}{3}} \left.\frac{\delta\rho}{\rho}\right|_0 \qquad (5)$$
$$T_{ta} = \left(\frac{3\pi}{4}\right)^{4/3} T_m$$

where $T_m$ is the temperature of the matter and $\left.\frac{\delta\rho}{\rho}\right|_0$ is the (hypothetical) linear density contrast at the present epoch.

The initial conditions (5) will be fixed by the power spectrum of the density perturbations that depends on the specific structure formation scenario.

## 3–Signal of a primordial cloud

As previously shown in MMT, proto–clouds at very high redshifts can give raise to a molecular signal if they are moving relatively to the Hubble flow. This effect is given by the resonant scattering of the CBR photons by the diatomic molecules in the proto–objects. In particular LiH molecule is responsible for the most important effect due to its high dipole moment.

The intensity of the effect is

$$\frac{\Delta I}{I_{CBR}} = (3 - \alpha_\nu)\frac{v_{pec,r}}{c}(1 - e^{-\tau_\nu}) \tag{6}$$

where $\alpha_\nu$ is the spectral index of the photons distribution ($\alpha_\nu = \frac{\nu}{I}\frac{dI}{d\nu}$), $\tau_\nu$ is the optical depth of the cloud calculated at the observational frequency, $v_{pec,r}$ is the component along the line of sight of the peculiar velocity of the cloud and $c$ is the light speed (see appendix).

The optical depth is given by the formula:

$$\tau = \int \sigma n_{LiH} dl \tag{7}$$

with

$$\sigma = \frac{\lambda_{ij}^3 A_{ji}}{4c}\frac{g_j}{g_i}\frac{\nu_{ij}}{\Delta\nu_D} \tag{8}$$

$$n_{LiH} = \Omega \omega_b n_{H,cr} \alpha_{LiH} n_{vJ} (1 + z)^3$$

where the notations mean: $\sigma$ = cross section averaged on the Doppler line broadening $\Delta\nu_D$; $A_{ji}$ = Einstein spontaneous emission coefficients; $n_{H,cr}$ = today numerical critic density of hydrogen ($\approx \frac{3}{4}n_{cr}$); $\alpha_{LiH}$ = numerical molecular abundance relative to the hydrogen; $n_{vJ}$ = fractional population of energetic level with quantum numbers $v$ and $J$.

The goal of this paper is to find out the expected intensity and line width of the molecular signal for a primordial cloud during the three different stages of evolution listed in the previous paragraph.

### 3.1–LiH abundance

LiH molecules are essentially formed by photoassociation through the process

$$\text{Li} + \text{H} \rightarrow \text{LiH} + \gamma \qquad (9)$$

when the temperature of the CBR became low enough to the photodissociation process to be effective. The LiH final abundance will be a function of the lithium produced by the primordial nucleosynthesis and of the efficiency of its conversion in LiH.

Lithium is observed in old disk and halo stars with an almost uniform abundance of $\frac{[\text{Li}]}{[\text{H}]} \sim 10^{-10}$ for $T_{eff} > 5500\ K$ (Spite & Spite 1993), and in young pop. I stars with a greater spread of values ranging from $10^{-10}$ to $2 \cdot 10^{-9}$ for the same $T_{eff}$ (Soderblom et al. 1993; Thorburn et al. 1993). The constancy of the oldest stars value is considered as a clue of its primordial origin or at least of the negligible lithium depletion in these stars (Deliyannis, Demarque & Kawaler 1990). Recently, however, lithium deficient halo stars were discovered (Thorburn 1992). Moreover a new class of stellar evolution models were developed: they take into account the effect of the loss of angular momentum in the lithium depletion, and explain the scattered values of population I stars and the *Spite plateau* of the halo stars at the same time. According to these models the primordial lithium abundance should be greater than $10^{-10}$, perhaps equal to the highest value measured in the pop. I stars (Pinsonneault, Deliyannis & Demarque 1992; Charbonnel, Vauclair & Zahn 1992; Zahn 1994). This is a part of the long standing "lithium problem" (Signore et al. 1994 and references therein).

The percentage of lithium converted to LiH is quite uncertain: the rate of radiative association and photodissociation are not known and this explain the large range of values: only 1% ÷ 10% of primordial lithium is converted for Dubrovich (1977) and Lepp & Shull (1984), while more recently Puy et al. (1993) suggested a value of 60% and Khersonskii & Lipovka (1993) found a total lithium conversion taking into account the temperature dependance of the chemical rates. The LiH primordial abundance can increase during the proto–cloud collapse (Puy & Signore 1995).

### 3.2–Amplitude and width of molecular lines

In order to know the probability to detect primordial molecular signals and to choose the best observational strategy (de Bernardis et al. 1993) we need to work out the intensity and the broadness of LiH lines. In this paragraph we will calculate the optical depth associated to a primordial cloud and the lineshape of the signal in the three different stages of evolution listed in the second paragraph. For the following we adopt a constant LiH abundance of $10^{-9}$; being the clouds always optically thin (except at the turn–around), the intensity scales linearly for other abundance values.

*Linear evolution:* As already described in MMT, in the case of a perturbation expanding with the Universe, the signal is spread on a range of frequencies ($\frac{\Delta \nu}{\nu} = \frac{\Delta z}{z}$) due to the different redshifts associated to different parts of the cloud. Each different strip, giving a signal at $\nu_{obs} = \frac{\nu_{mol}}{1+z_{strip}}$ has an associated optical depth $\tau_{\nu_{obs}}$. The shape of the molecular line follows the distribution of $v_{pec,r}$ along the cloud; assuming a cloud of dimension $D$ moving with the same $v_{pec}$, the line width is given by:

$$\frac{\Delta \nu}{\nu} = D \frac{H_0}{c} (1+z) \sqrt{1+\Omega z} = 5.77 \cdot 10^{-4} \, m_{12}^{1/3} \sqrt{1+z} \qquad (10)$$

where $m_{12} = \frac{M}{10^{12} M_\odot}$.

*Turn–around:* When the cloud stops to follow the Hubble flow, each part contributes to the same frequency; this is the case where the line amplitude, i.e. the optical depth, is the highest and its width is the thinnest due only to the Doppler broadening. For a Maxwellian distribution of the velocities we have:

$$\frac{\Delta\nu}{\nu} = \frac{2}{c}\sqrt{\frac{2\ln 2 k_B T}{m_a}} = 7.16 \cdot 10^{-7}\sqrt{\frac{T}{A}} = 2.541 \cdot 10^{-7}\sqrt{T} \qquad (11)$$

where $m_a$ and $A$ are the atomic mass and the atomic number of the molecule and the last equality is effective for LiH.

*Non–linear collapse*: In a more advanced collapsing phase we have to take into account two kinds of velocities: the peculiar and the infall velocity. The presence of the second one has as a consequence that different strips of the cloud contribute to the signal at different frequencies.

In order to evaluate the shape and the amplitude of the line we assume a radial profile for the density and for the infall velocity of the cloud. We take:

$$\begin{aligned} \rho(r) &= \tilde{\rho}\cos(kr) \\ v_c(r) &= \tilde{v}_c \sin(kr) \end{aligned} \qquad (12)$$

where $r$ is the radial coordinate along the cloud and $\tilde{v}_c$ is the infall velocity of the external shell. From the matching conditions we get $k = \frac{\pi}{2R}$ with $R =$ radius of the cloud and if $\bar{\rho}$ is the mean density we get $\tilde{\rho} = \frac{\pi^3 \bar{\rho}}{6(\pi^2-8)}$.

Due to their different infall velocity, different parts of the clouds contribute to the signal at different frequencies. We have to search for surfaces characterized by an equal projection of the infall velocity along the line of sight; these surfaces will emit a signal at a frequency

$$\nu_s = \frac{\nu_0}{1+z}\left[1 - \frac{\tilde{v}_c}{c}\sin(kr)\cos\theta\right] \qquad (13)$$

whose amplitude is:

$$\left.\frac{\Delta I}{I_{CBR}}\right|_s = (3-\alpha_s)\left[\left(\vec{\beta}_c\cdot\vec{n}\right)_s + \vec{\beta}_{pec}\cdot\vec{n}\right]\left(1 - e^{-\tau_s}\right) \qquad (14)$$

where $\theta$ is the angle between the infall velocity and the line of sight, $\vec{n}$ is the line of sight unit vector and the subscript $s$ refers to isovelocity surfaces (see figure 1).

In figure 2 is shown the lineshape for different values of $\tilde{v}_c$ as regards $v_{pec}$. For $\vec{\beta}_{pec} \cdot \vec{n} = 0$ the line is characterized by an absorption and an emission peak that turn into a single peak line (in emission or absorption depending on the peculiar velocity direction) for $\vec{\beta}_{pec} \cdot \vec{n} > \tilde{v}_c/c$. From equation (13), the line width is:

$$\frac{\Delta \nu}{\nu} \approx 2 \frac{\tilde{v}_c}{c} \qquad (15).$$

It is worthwhile to stress that the non–linear collapse is the only stage of the proto–cloud evolution where resonant scattering lines can exist even with a zero peculiar velocity component along the line of sight.

### 3.3–The peculiar velocity

To have an idea of the expected magnitude of this signal we have to study the distribution of the peculiar velocities of these proto–objects. We can adopt a statistical approach. Let's consider the equation of continuity in the wavenumber space:

$$\dot{\delta}_k + i \frac{\vec{k}}{a} \cdot \vec{v}_k = 0 \qquad (16)$$

$\delta_k$ is the $k$–th component of the expansion of $\frac{\delta\rho}{\rho}$, $\vec{k}$ is the corresponding wavenumber, $\vec{v}_k$ is its velocity and $a$ is the scale factor of the universe.

Since $\delta_k \propto t^{2/3}$ (for $\Omega = 1$) and $\vec{\nabla} \wedge \vec{v} = \vec{0}$ it is easy to obtain:

$$|\vec{v}_k| = a(t)H(t)\frac{\delta_k(t)}{k} \propto a(t)^{1/2} \propto \frac{1}{\sqrt{1+z}} \qquad (17)$$

The rms value of the peculiar velocity averaged over a spherical region of radius $R$ is given by

$$v_{rms}^2 = a^2(t)H^2(t) \int \frac{P(k,t)}{2\pi^2} W^2(kR) dk \qquad (18)$$

where $W(kR)$ is the window function.

### 4.1–Observability of a primordial cloud: the CDM model

Considering the CDM model, we take the associated power spectrum for the density perturbations (Peebles 1983; Davis et al. 1985):

$$P(k) = \frac{Ak}{(1 + \beta k + \omega k^{3/2} + \gamma k^2)^2} \qquad (19)$$

with $\beta = 1.7(\Omega_0 h^2)^{-1}\ Mpc$, $\omega = 9(\Omega_0 h^2)^{-3/2}\ Mpc^{3/2}$, $\gamma = (\Omega_0 h^2)^{-2}\ Mpc^2$ and $A = (24h^{-1}\ Mpc)^4$ (post–COBE normalization) and where $k$ is the wave number. In figure 3 we plot the spectrum given by equation (19) and the component along the line of sight of $v_{rms}^2$ associated to this spectrum (see equation [18]).

We will consider clouds with four different masses $M = 10^9, 10^{10}, 10^{11}, 10^{12}\ M_\odot$, which attempt the turn–around respectively at $z_{ta} \sim 12, 8, 5, 3$; at this stage their angular dimensions are respectively $4.6''$, $10.9''$, $26.3''$, $67.1''$. However these turn–around redshifts are just indicative (they can be shifted by some unity) either due to the statistical feature of the spectrum (19) or to the uncertainties existing in the CDM theory of structure formation.

Results of the evolution of different physical quantities for the clouds during their non linear collapse are collected in figure 4 and table 2. With the adopted initial molecular hydrogen abundance the cooling function does not affect significantly the collapse, which is quasi–adiabatic.

In figure 5 we plot the evolution of the molecular signal throughout the different stages of the primordial clouds evolution. The dashed lines mark the transition region between the linear region and the turn–around, that is for $1 \leq \frac{\delta \rho}{\rho} \leq 4.55$, where linear approximation is no more effective.

During the linear phase the signal from the cloud is quite low but, as described in MMT, it is possible to obtain values five times greater than those in figure 5 for transitions with higher values of $J$. Anyway the main problem arises from an experimental point of view. If we want to use a radiotelescope to detect these signals we have to take into account the problem of line width. For example, taking an observational condition $\frac{\Delta \nu}{\nu} < 10^{-3}$ the equation (10) leads to $m_{12}^2 (1+z)^3 < 27.1$. For clouds not satisfying this condition the line width gives problems for the baseline of the spectrum, because it is necessary to superimpose different spectra with a significant loss of sensibility.

At turn–around time the signal is higher but the line is too thin (see formula [11]) and so it is not possible to detect it with backends normally available at radiotelescopes.

The collapsing phase is the most favorable one for both the line width and the signal amplitude. The signal intensity goes down after the turn–around because of the spread of the line due to the fact that only the part of the cloud satisfying condition (13) will give a contribution to the signal. After reaching a minimum for an evolutionary stage of $t/t_{ff} \sim 0.5$, the $J = 0$ line intensity grows up taking advantage of the improvement of the peculiar velocity and of the cloud density. The increase for the $J = 1$ line is delayed by the depopulation towards the fundamental level. This behaviour is more marked at higher masses because lower characteristic temperatures are associated with the smaller redshifts at which the clouds collapse.

Only at the higher redshifts associated with a $10^9 \, M_\odot$ cloud collapse, the $J = 1$ line gives the highest signal due to the $(3 - \alpha)$ factor in formula (6). For higher values of $J$ the population drops down and this decrease is no more compensated by the previous factor. For the same reason the $J = 0$ line is the strongest one for the clouds of higher masses, that collapse at lower redshifts.

To search for primordial molecules signal, it is essential to be able to detect simultaneously two rotational lines: two signals at the right redshifted frequencies would be the safest way to confirm a detection. If we choose, as an observational criterion, a line width $\frac{\Delta \nu}{\nu}$ between $10^{-5}$ and $10^{-3}$ and an intensity $\frac{\Delta I}{I_{cbr}}$ greater than $3 \cdot 10^{-4}$ we can identify two different favourable epochs in the history of a primordial cloud evolution. The first one is at the beginning of the non linear collapse phase between $t/t_{ff} \sim 0.05 \div 0.1$ and $t/t_{ff} \sim 0.1 \div 0.2$.

While for the $J = 0$ line the signal improves for growing masses, the range of observability for both the lines at the same times is reduced due to the drop of the $J = 1$ line intensity. Due to this drop, the second epoch for $t/t_{ff} \geq 0.8 \div 0.9$ is available only for masses smaller than $10^{11}\, M_\odot$. In any case our simple model of spherical collapse is no more correct for such advanced phases of collapse.

In the first epoch the two redshifted lines are observable for frequencies between $30 \div 250\, GHz$ and the angular dimensions range from $3''$ to $1'$.

### 4.2–Observability of a primordial cloud: the general case

In the previous subsection we focused our attention on a particular cosmological model. In principle it is possible to extend our analysis of the first steps of collapse to any model. For example in a top–down scenario (such as HDM models) the large–scale structures form very early and $z_{ta}$ is, in general, greater.

We calculated the signal for clouds of mass ranging from $10^8$ to $10^{13}\, M_\odot$ with $z_{ta} = 10, 20, 30, 40, 50, 70$, and $\omega_B = 0.1$, without referring to any particular structure formation scenario. Results are collected in figures 6 and 7.

At the turn–around, the smaller is the redshift $z_{ta}$, the higher is the velocity $v_{pec}$ and therefore the higher is the signal. The signal is almost mass indipendent due to the fact that the $J = 0, 1$ lines are optically thick at the turn–around.

During the collapse the $J = 1$ line is the brightest for all the values of $z_{ta}$ greater than 10 confirming the explanation given in the previous paragraph. Comparing to the CDM case, we have always the same two favourite regions for detection, but they are delimited by the fundamental line intensity. The observational frequencies are shifted towards lower values.

Studying the signal dependance from the mass of the cloud and $z_{ta}$ during the collapse, we can write equation (6) as

$$\frac{\Delta I}{I_{CBR}} \propto (3 - \alpha)[\beta_c + \beta_{pec}]\left[1 - \exp\left(-C\frac{M}{\beta_c R^2}n_{vJ}f_{ls}\right)\right] \qquad (20)$$

where $f_{ls}$ is a correction factor depending on the line shape: $f_{ls}$ tends to 1 when the peak is near the central frequency of the line ($\beta_{pec} \gg \beta_c$) because the highest contribution to the signal comes from the equatorial region of the cloud, while $f_{ls}$ tends to 0 when the peak is displaced from the central frequency ($\beta_{pec} \ll \beta_c$), the highest contribution coming from high latitudes regions of the cloud where the density is smaller.

Expliciting the $(z, M)$ dependance of the terms in formula (20) and assuming for $\beta_c$ the same behaviour of the adiabatic collapse, we can distinguish four different regimes:

$\tau \geq 1$

$$\frac{\Delta I}{I_{CBR}} \propto \frac{(3 - \alpha)}{\sqrt{1 + z}} \qquad \beta_c < \beta_{pec} \quad (21)$$

$$\frac{\Delta I}{I_{CBR}} \propto (3 - \alpha)\sqrt{1 + z_{ta}}M^{1/3} \qquad \beta_c > \beta_{pec} \quad (22)$$

$\tau \ll 1$

$$\frac{\Delta I}{I_{CBR}} \propto (3 - \alpha)n_{vJ}f_{ls}(1 + z_{ta})\sqrt{\frac{1 + z_{ta}}{1 + z}} \qquad \beta_c < \beta_{pec} \quad (23)$$

$$\frac{\Delta I}{I_{CBR}} \propto (3 - \alpha)n_{vJ}f_{ls}(1 + z_{ta})^2 M^{1/3} \qquad \beta_c > \beta_{pec} \quad (24)$$

Except for the very early phases of the collapse ($t/t_{ff} \leq 0.1$), the primordial clouds are always optically thin.

Referring to figure 6, the signal slightly depends on $z_{ta}$ for masses smaller than $10^{10} M_\odot$: at a fixed $t/t_{ff}$ the linear dependance of formula (23) is smoothed by the factor $(3-\alpha)n_{vJ}$ that decreases at high redshifts. Physically the column density, proportional to $(1+z_{ta})^2$, is partly compensated at high redshift by the spread of the line (i.e. the contribution of the molecules is shared by a larger range of frequencies) and the low values of the peculiar velocity.

For higher masses, the $z_{ta}$ dependance is enhanced by the passage to the $\beta_c > \beta_{pec}$ regime, starting with high values of $M$, $z_{ta}$ and $t/t_{ff}$; in this case the spread of the line is balanced by the increase of the total peculiar velocity $(\beta_{pec} + \beta_c)$ and the signal intensity follows the column density dependance.

In figure 7 we can see the signal independence from the mass of the cloud till it remains in the regime dominated by the peculiar velocity, that is at low redshifts, for low masses and during the initial phases of the collapse (as suggested by formulas [21] and [23]).

## 5–Conclusions

Starting from a simple model of collapse for primordial clouds, we studied the possible signal due to the resonant elastic scattering on LiH molecules in the framework of a CDM scenario and in a more general situation for various values of the mass $M$ and the turn–around redshift $z_{ta}$ of the proto–clouds.

We found that the most favourable epoch for the production of a molecular signal is the first phase of the non linear collapse, soon after the turn–around time.

The molecular line intensity $\frac{\Delta I}{I_{CBR}}$ and width $\frac{\Delta \nu}{\nu}$ are not strongly dependent of $M$ and $z_{ta}$, except in the *double peak line regime*, i.e. when the infall velocity is higher than the peculiar one ($\beta_c > \beta_{pec}$). Conversely the observational frequency $\nu_{obs}$ and the angular size $\theta$ depend on $M$ and $z_{ta}$. Therefore the detection of - or even the estimation of an upper limit on - these molecular signals will strongly constrain the various structure formation models.

Despite the weakness of the expected signals, they have very important characteristics that can be useful for detection: the molecular character of the effect gives to the expected signal the peculiarity of having spectral features; therefore the local continuum backgrounds will not affect the measurements.

## Acknowledgments

This research program was partly supported by the EEC network CHRX-CT920079 on "The CMB radiation measurements and interpretations". One of the authors (R. M.) is grateful to the European Space Agency for the financial support.

# Appendix

Let us assume that a cloud at redshift $z$ is moving with a peculiar velocity $\beta_{pec} = \frac{v_{pec}}{c}$ with an angle $\theta_0$ respect to the line of sight. An observer O is receiving photons from the cloud within its solid angle $\Omega_{obs}$: if $\nu_0$ is the resonant line rest frequency, the observer has to tune his receiver to $\nu_{obs} = \frac{1}{1+z}\nu_0 \left(1 + \beta_{pec}\cos\theta_0\right)$.

Photons arriving from the cloud into the field of view of the detector are composed by two terms:

*a)* First of all there are photons passing through the cloud without being scattered: they are a fraction $e^{-\tau}$ of the total CBR photons crossing the cloud within the solid angle $\Omega_{obs}$ and with the frequency $\nu' = \nu_0\left(1 + \beta_{pec}\cos\theta_0\right)$

$$N_{ns} = N_{BB}\left(\nu', T'\right) e^{-\tau} \Omega_{obs}$$

where $T' = 2.74\left(1 + z_{cloud}\right)$ is the CBR temperature at the cloud redshift and $N_{BB}$ is the black body photon number density.

*b)* Secondly, we have to consider CBR photons coming from every direction and scattered by the cloud towards the observer within the solid angle $\Omega_{obs}$. We schematize the resonant scattering as an absorption followed by an isotropic emission in the cloud reference frame. The number of the absorbed photons averaged on the solid angle and calculated in the cloud frame is:

$$\begin{aligned}
N_{\Omega,abs} &= \frac{1 - e^{-\tau}}{4\pi} \int N_{BB}\left(\nu_0, T'\frac{1 + \beta_{pec}\cos\theta'}{\sqrt{1 - \beta_{pec}^2}}\right) d\Omega \\
&= \frac{1 - e^{-\tau}}{2} \left[\int N_{BB}\left(\nu_0, T'\right) \sin\theta' d\theta' + \right. \\
&\quad \left. + \int T'\beta_{pec}\cos\theta' \left.\frac{\partial N_{BB}\left(\nu_0, T'\right)}{\partial T}\right|_{T'} \sin\theta' d\theta'\right] \\
&= \left(1 - e^{-\tau}\right) N_{BB}\left(\nu_0, T'\right)
\end{aligned}$$

where we consider the Doppler effect on the CBR temperature in the cloud frame and we neglect terms at second order in $\beta_{pec}$ being always $\beta_{pec} \ll 1$.

These photons are emitted isotropically in the cloud frame, so we have to multiply $N_{\Omega,abs}$ for the receiver solid angle *in the cloud frame*, that is

$$\Omega_{cloud} = \frac{1 - \beta_{pec}^2}{1 - 2\beta_{pec}\cos\theta_0}\Omega_{obs} \simeq (1 + 2\beta_{pec}\cos\theta_0)\,\Omega_{obs}$$

The differential signal will be

$$\frac{N_{cloud} - N_{sky}}{N_{sky}} = \frac{[N_{ns} + N_{\Omega,abs}\Omega_{cloud}] - N_{BB}(\nu',T')\,\Omega_{obs}}{N_{BB}(\nu',T')\,\Omega_{obs}}$$
$$= \frac{N_{BB}(\nu',T')\,e^{-\tau} + (1 - e^{-\tau})\,N_{BB}(\nu_0,T')(1 + 2\beta_{pec}\cos\theta_0) - N_{BB}(\nu',T')}{N_{BB}(\nu',T')}$$

Using $\nu_0 = \nu'(1 - \beta_{pec}\cos\theta_0)$ and developing $N_{BB}(\nu_0,T')$ at first order in $\nu'$, we have

$$\begin{aligned}\frac{\Delta N}{N} &= e^{-\tau} + (1 - e^{-\tau})\left[1 - \beta_{pec}\cos\theta_0 \frac{\nu'}{N_{BB}(\nu',T')}\left.\frac{\partial N_{BB}(\nu',T')}{\partial \nu}\right|_{\nu'}\right] \times \\ &\quad \times (1 + 2\beta_{pec}\cos\theta_0) - 1 \\ &= (1 - e^{-\tau})\left[(1 - \beta_{pec}\cos\theta_0\,\alpha'_N)(1 + 2\beta_{pec}\cos\theta_0) - 1\right] \\ &= (1 - e^{-\tau})\,\beta_{pec}\cos\theta_0\,(2 - \alpha'_N)\end{aligned}$$

where $\alpha'_N = \frac{\nu'}{N_{BB}(\nu',T')}\left.\frac{\partial N_{BB}(\nu',T')}{\partial \nu}\right|_{\nu'}$ is the spectral index for the photon number.

The term $\beta_{pec}\cos\theta_0\,\alpha'_N$ arises from the fact that the photons processed by the cloud comes in average from a different frequency respect to the observational one, while the term $2\beta_{pec}\cos\theta_0$ is due to the non isotropic emission from the cloud in the observer frame.

Using the equality

$$\alpha'_N = \frac{\nu'}{N_{BB}(\nu',T')} \left.\frac{\partial N_{BB}(\nu',T')}{\partial \nu}\right|_{\nu'} =$$
$$= \frac{\nu_{obs}}{N_{BB}(\nu_{obs},2.74)} \left.\frac{\partial N_{BB}(\nu_{obs},2.74)}{\partial \nu}\right|_{\nu_{obs}} = \alpha_N$$

and the relation between the spectral indexes $\alpha_N = \alpha_\nu - 1$ where $\alpha_\nu = \frac{\nu}{I}\frac{\partial I}{\partial \nu}$, we have

$$\frac{\Delta N}{N} = \frac{\Delta I}{I_{CBR}} = \left(1 - e^{-\tau}\right) \beta_{pec} \cos\theta_0 \left(3 - \alpha_\nu\right)$$

where the spectral index is calculated for the black body at $2.74\,K$ and at the observer frequency.

**Table 1**

| Reactions | Rates | References |
|---|---|---|
| $H + e \rightarrow H^- + h\nu$ | $k_1 = 1.83 \cdot 10^{-18} T_{mat}^{0.8799}$ | Hutchins 1976 |
| $H + H^- \rightarrow H_2 + e$ | $k_2 = 1.35 \cdot 10^{-9}$ | Hutchins 1976 |
| $H^+ + e \rightarrow H + h\nu$ | $k_3 = 1.88 \cdot 10^{-10} T_{mat}^{-0.644}$ | Hutchins 1976 |
| $H^- + h\nu \rightarrow H + e$ | $k_4 = 0.278\, T_{rad}^2 \exp(-8.750/T_{rad})$ | Hirasawa 1969 |

**Table 2**

| Mass $(M_\odot)$ | $z_{ta}$ | $r_{ta}$ $(kpc)$ | $\theta_{ta}$ | $T_{ta}$ $(K)$ | $T_{t_{ff}}$ $(K)$ | $v_{t_{ff}}$ $(km/s)$ |
|---|---|---|---|---|---|---|
| $10^9$ | 12 | 5.0 | 4.6″ | 12.73 | 101.09 | 67.23 |
| $10^{10}$ | 8 | 15.6 | 10.9″ | 6.47 | 51.49 | 120.71 |
| $10^{11}$ | 5 | 50.3 | 26.3″ | 3.07 | 24.42 | 212.39 |
| $10^{12}$ | 3 | 162.6 | 67.1″ | 1.46 | 11.58 | 373.62 |

# References

Bond J. R., Carr B. J. & Hogan C. J. 1991, ApJ, 367, 420

Charbonnel C., Vauclair S. & Zahn, J.-P. 1992, A&A, 255, 191

Cole S. & Kaiser N. 1988, MNRAS, 233, 637

Davis M., Efstathiou G., Frenk C. S. & White S.D.M. 1985, ApJ, 292, 371

De Araujo J. C. N. & Opher R. 1988, MNRAS, 231, 923

de Bernardis P., Dubrovich V., Encrenaz P., Maoli R., Masi S., Mastrantonio G., Melchiorri B., Melchiorri F., Signore M. & Tanzilli P. E. 1993, A&A, 269, 1

Deliyannis C. P., Demarque P. & Kawaler S. D. 1990, ApJS, 73, 21

Dubrovich V. K. 1977, Sov. Astron. Lett., 3, 128

Dubrovich V. K. 1993, Sov. Astron. Lett., 19, 132

Gunn J. E. & Gott III R. J. 1972, ApJ, 176, 1

Hirasawa T. 1969, Prog. Theor. Phys., 42, 523

Hu W., Scott D. & Silk J. 1993, Phys. Rev. D, 49, 648

Hutchins J. B. 1976, ApJ, 205, 103

Khersonskii V. K. & Lipovka A. A. 1993, Astrofizicheskie Issledovaniya, 36, 88

Lahav O. 1986, MNRAS, 220, 259

Lepp S. & Shull M. J. 1983, ApJ, 270, 578

Lepp S. & Shull M. J. 1984, ApJ, 280, 465

Maoli R., Melchiorri F. & Tosti D. 1994, ApJ, 425, 372 (MMT)

Martìnez-Gonzàlez E., Sanz J. L. & Silk J. 1990, ApJ, 355, L5

Ostriker J. P. & Vishniac E. T. 1986, ApJ, 306, L51

Padmanabhan T. 1993, Structure formation in the universe (Cambridge University Press)

Palla F., Salpeter E. E. & Stahler S. W. 1983, ApJ, 271, 632

Palla F., Galli D. & Silk J. 1995, ApJ, submitted

Peebles P. J. E. 1983, ApJ, 263, L1

Peebles P. J. E. 1993, Principles of Physical Cosmology (New York: Princeton University Press)

Pinsonneault M. H., Deliyannis P. D. & Demarque P. 1992, ApJS, 78, 179

Puy D., Alecian G., Le Bourlot J., Leorat J., Pineau des Forets G. 1993, A&A, 267, 337

Puy D. & Signore M., 1995, A&A, submitted

Rees M. J. & Sciama D. W. 1968, Nature, 517, 611

Schaeffer R. & Silk J. 1988, ApJ, 333, 509

Signore M., Vedrenne G., de Bernardis P., Dubrovich V., Encrenaz P., Maoli R., Masi S., Mastrantonio G., Melchiorri B., Melchiorri F. & Tanzilli P. E. 1994, ApJS, 92, 535

Silk J. 1983, MNRAS, 205, 705

Soderblom D. R., Burton F. J., Balachandran S., Stauffer J. R., Duncan D. K., Fedele S. B., Hudon J. D. 1993, AJ, 106, 1059

Spite F. & Spite M. 1993, A&A, 279, L9

Subramaniam K.& Padmanabhan, T. 1993, MNRAS, 265, 101

Thorburn J. A. 1992, ApJ, 399, L83 Thorburn J. A., Hobbs L. M., Deliyannis C. P., Pinsonneault M. H. 1993, ApJ, 415, 150

Vishniac E. T. 1987, ApJ, 322, 597

Zahn J.-P. 1994, A&A, 288, 829

**Figure captions**

**Fig. 1**: Slice of a spherical cloud during the collapsing phase; the plotted lines are a section of the surfaces contributing to the signal at the same observational frequency. The numbers are the values of the $v_c(r)/\tilde{v}_c$ ratio (see formula [12]).

**Fig. 2**: Different shapes in arbitrary units of the signal for different ratio between the peculiar and the infall velocities with $v_c = 30\,Km/s$. The molecular lines are supposed to be optically thin and refer to $\beta_{pec}/\beta_c = 0, 0.33, 1, 5$ respectively from the top to the bottom and from the left to the right. It is more likely to have the double peak feature for high mass primordial clouds at high redshift and in an advanced phase of collapse.

**Fig. 3a**: Primordial power spectrum associated to a CDM model (formula [19]).

**Fig. 3b**: Rms value of the peculiar velocity obtained from the formula (18) for the CDM spectrum.

**Fig. 4**: Normalized values of some fundamental quantities during the collapsing phase of a cloud. The normalization factors for clouds of masses between $10^9$ and $10^{12}\,M_\odot$ are reported in table 2.

**Fig. 5**: On the top the intensity and line width (long dashed line) are reported for the $J = 0, 1$ lines of primordial clouds of masses between $10^9$ and $10^{12}\,M_\odot$ in the CDM scenario. The upper scale is the observational frequency for the two lines. The collapse phase $t/t_{ff}$ in steps of 0.1 is also reported in the figure. In the bottom we plot the angular dimension of the cloud. The short dashed line indicates the $\frac{\delta\rho}{\rho} \geq 1$ region where the linear approximation is no more effective. The molecular signal passes from a low intensity broad line in the linear phase through a high intensity thin line at the turn–around to an intermediate intensity line of growing broadness during the non linear collapse.

**Fig. 6**: Line intensity, line width, peculiar and infall velocity as a function of the collapse phase $t/t_{ff}$, for clouds with $z_{ta} = 10, 20, 30, 40, 50, 70$ at fixed $J$ and fixed mass. The crossing between the $\beta_{pec}$ and the $\beta_c$ lines determines the passage to the *double peak line* regime: a $10^8 \, M_\odot$ cloud has always $\beta_{pec} > \beta_c$, while for example a $10^{13} \, M_\odot$ cloud moves to the $\beta_c > \beta_{pec}$ regime at a very early phase of the collapse for all the values of $z_{ta}$.

**Fig. 7**: The $J = 0, 1$ lines intensity as a function of the collapse phase $t/t_{ff}$ for clouds of mass $10^9 \div 10^{12} \, M_\odot$ at fixed turn–around redshifts.

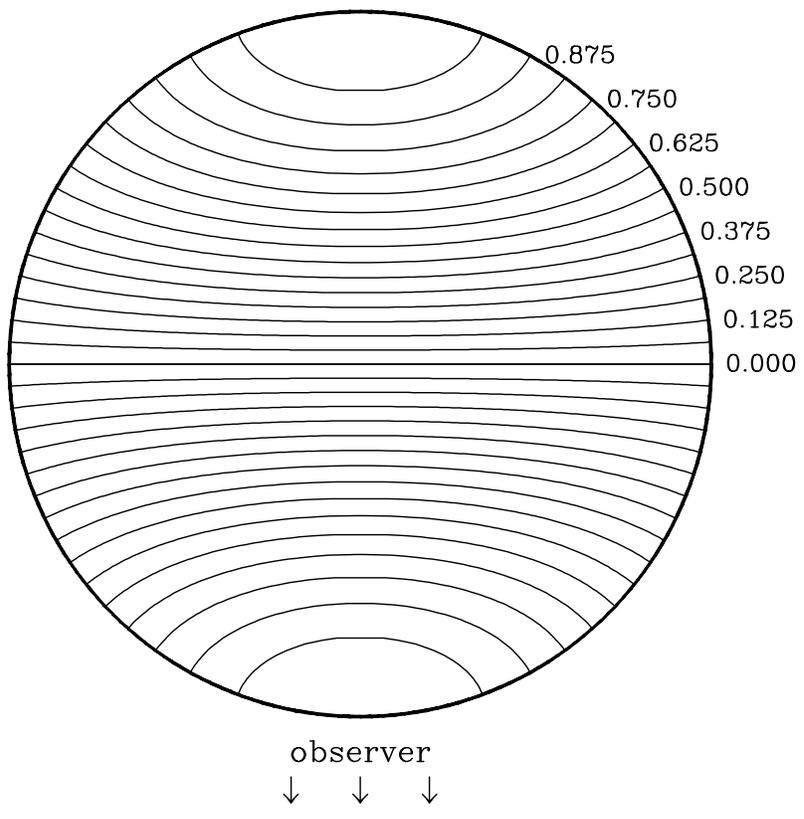

Fig. 1

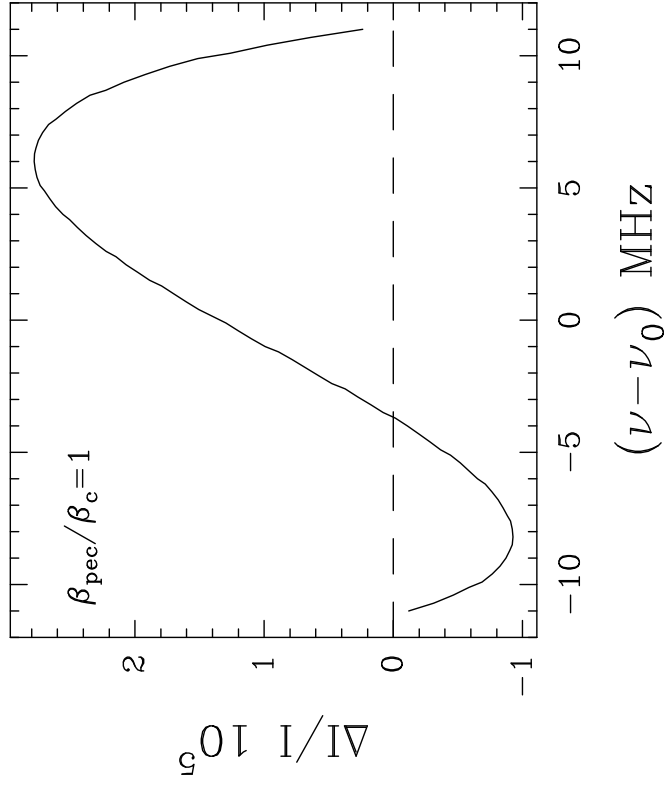
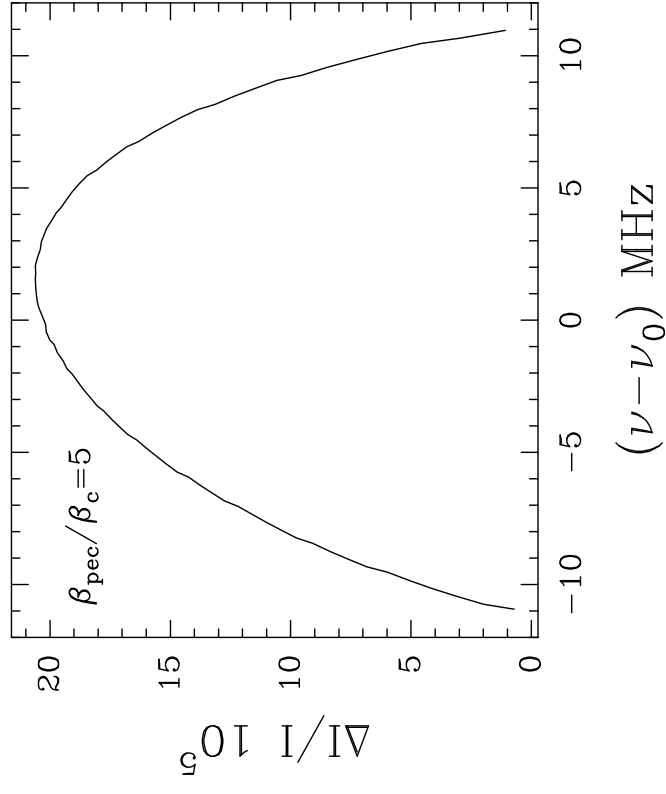
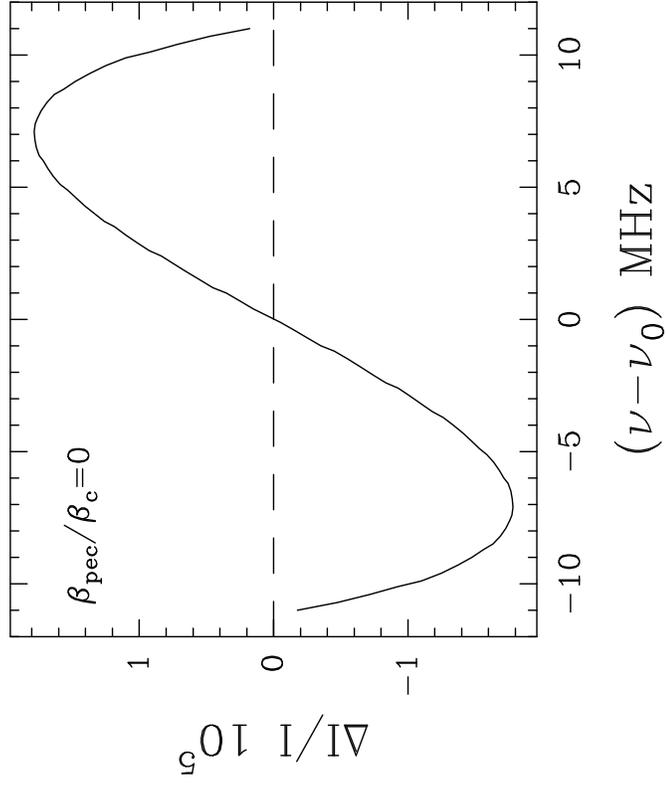
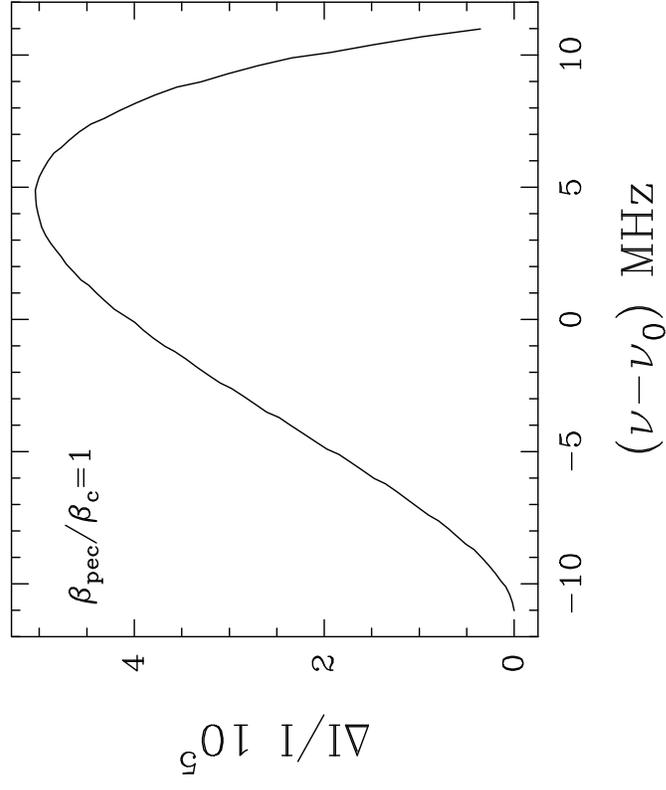

Fig.

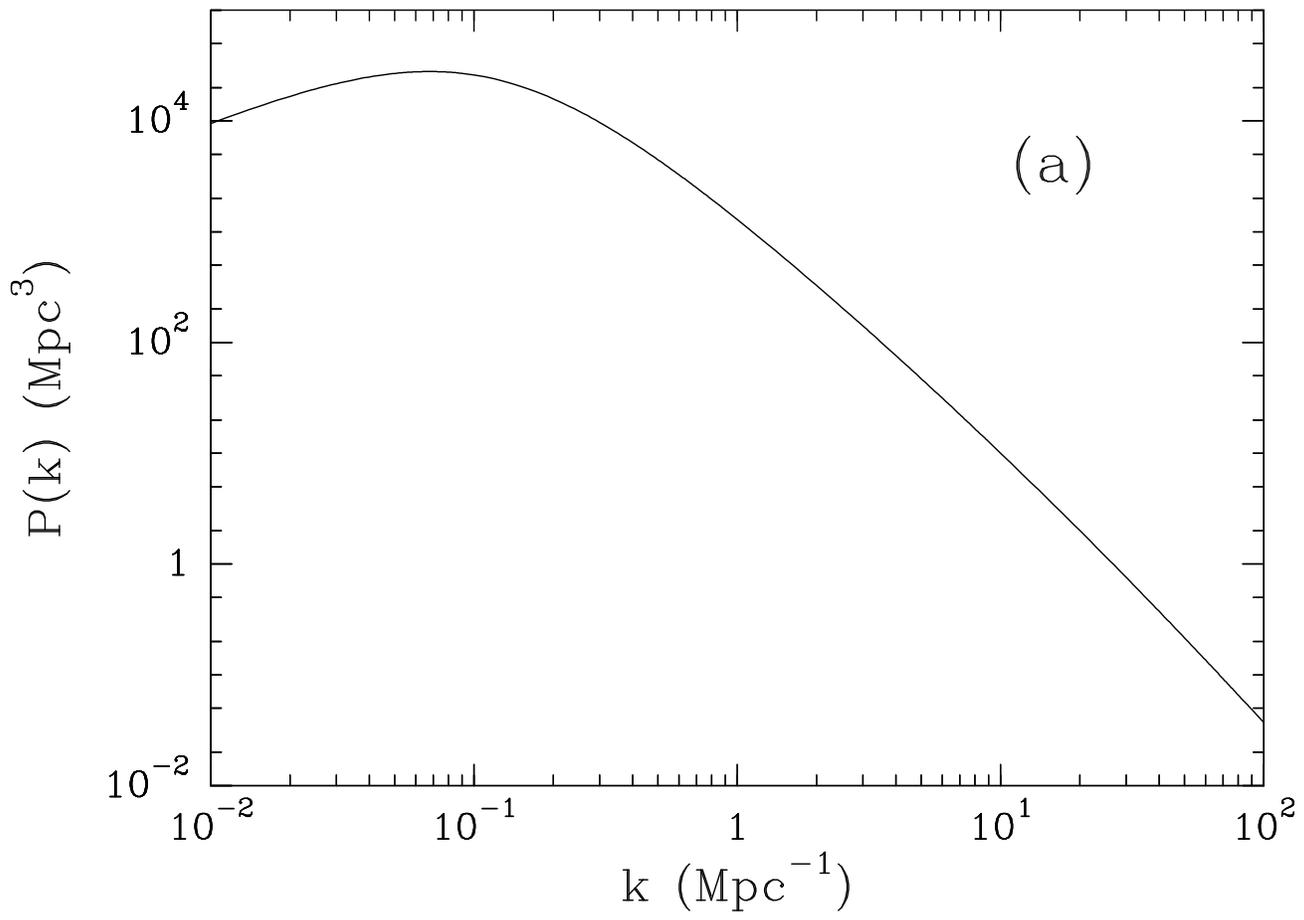

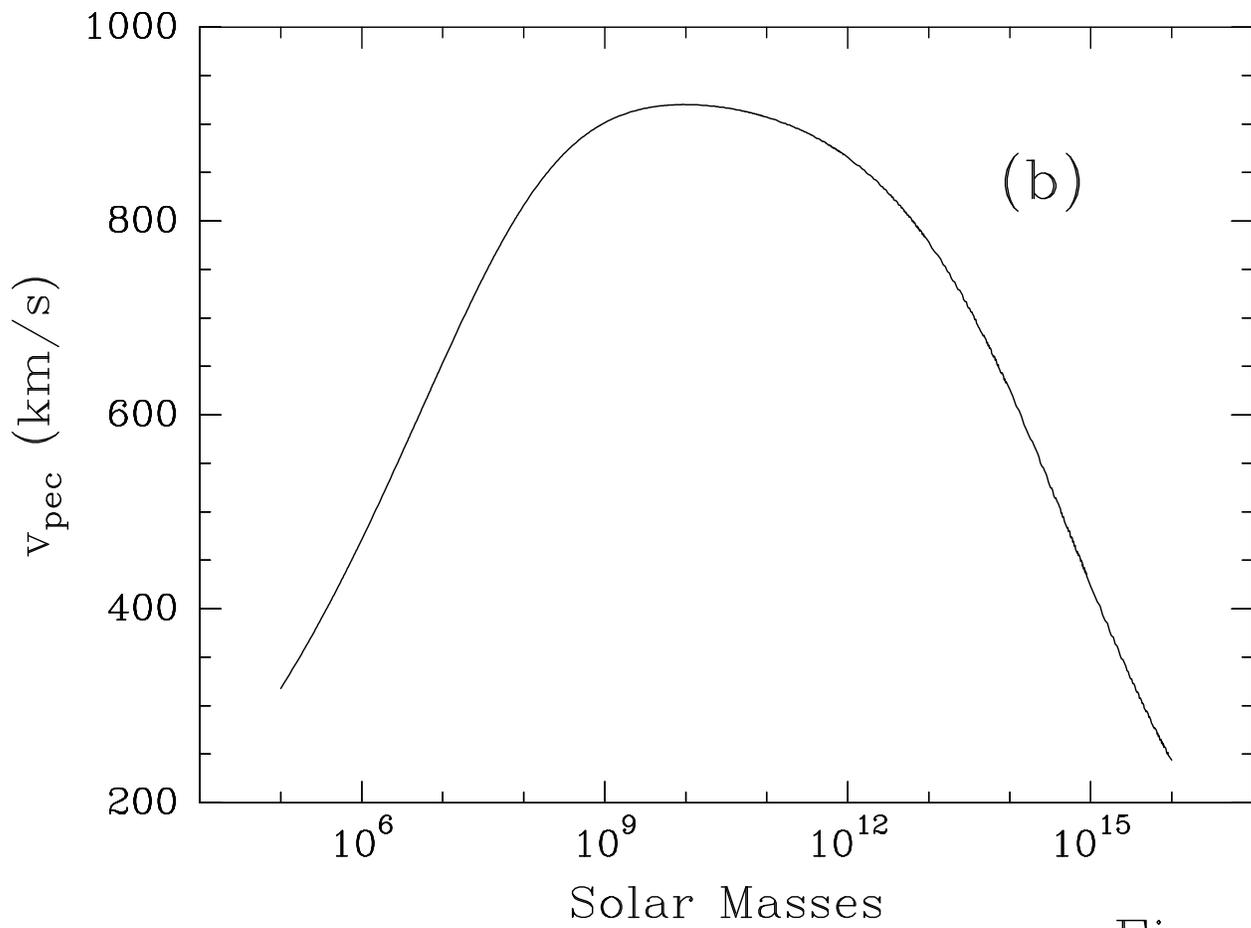

Fig. 3

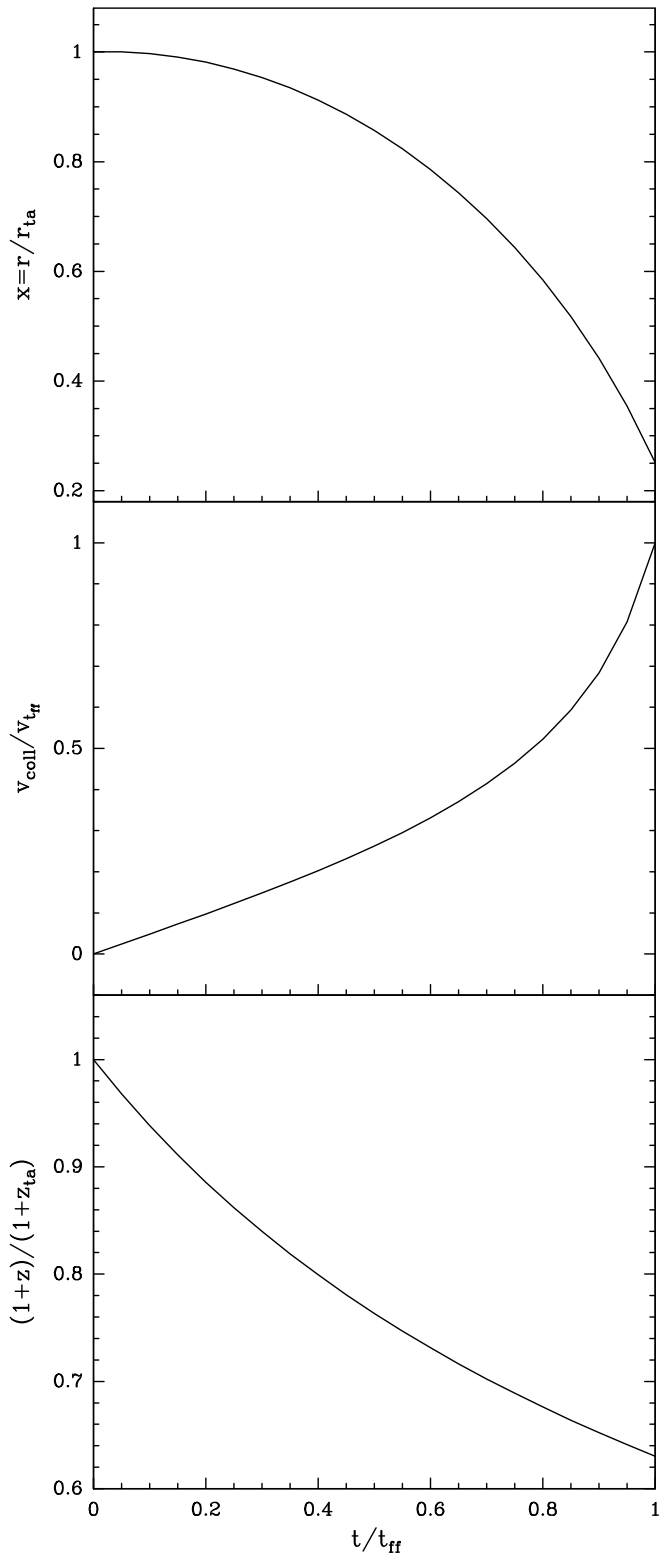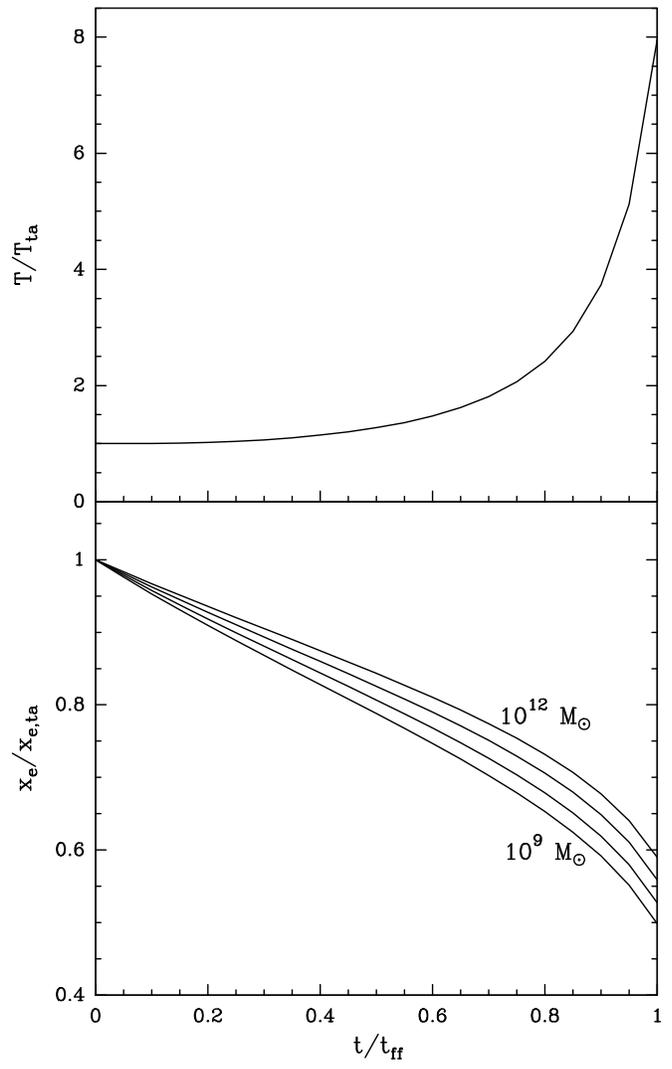

Fig. 4

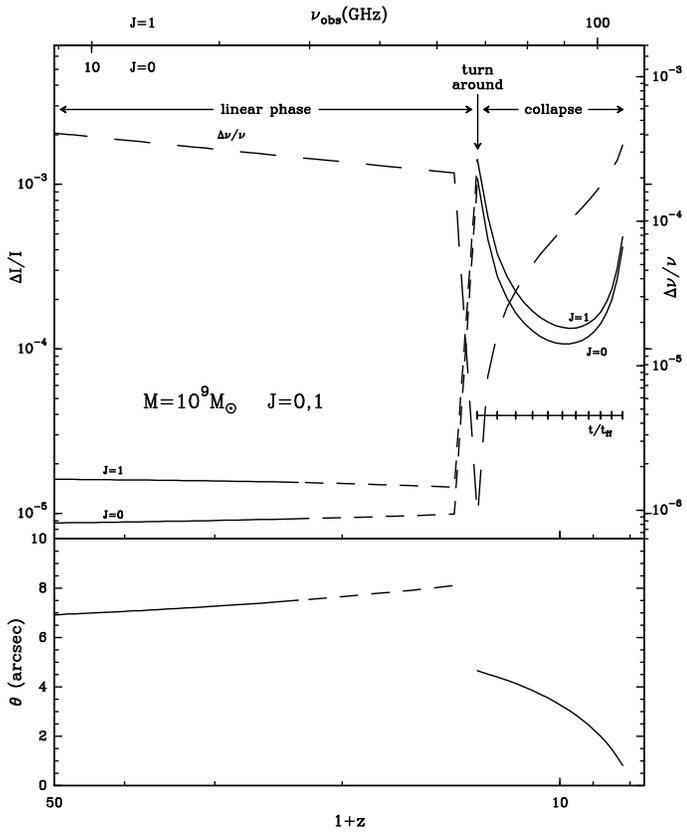
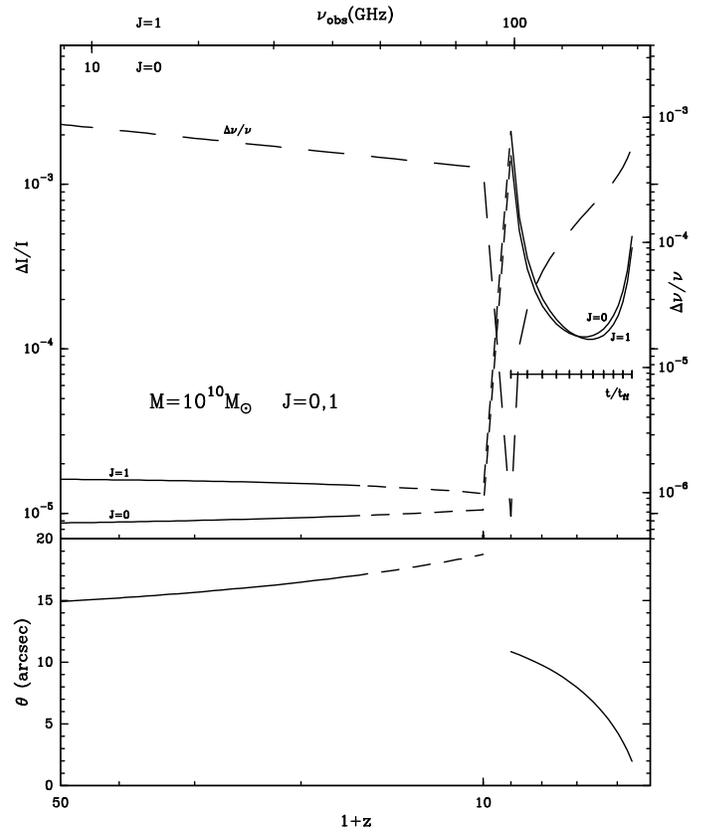
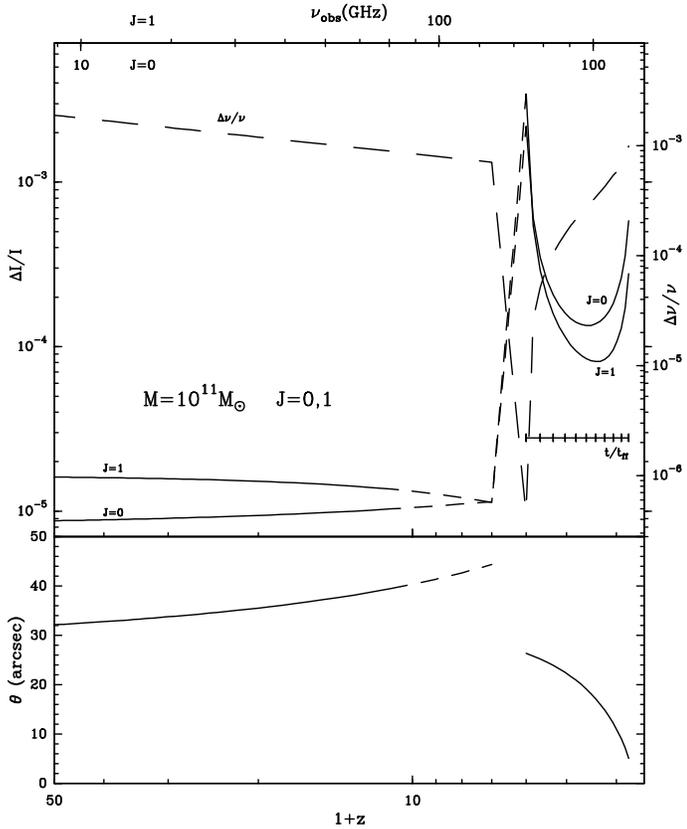
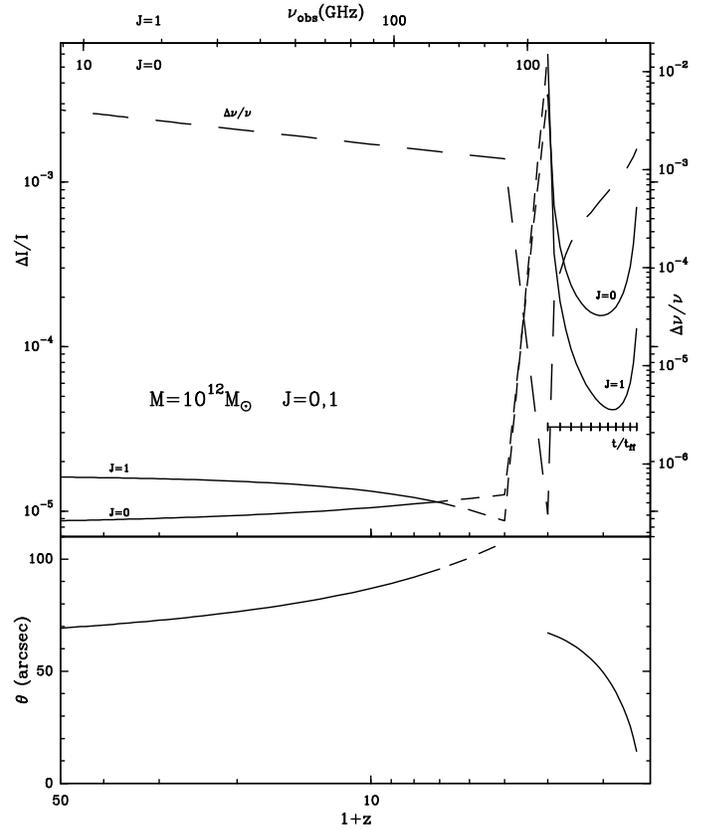

Fig. 5

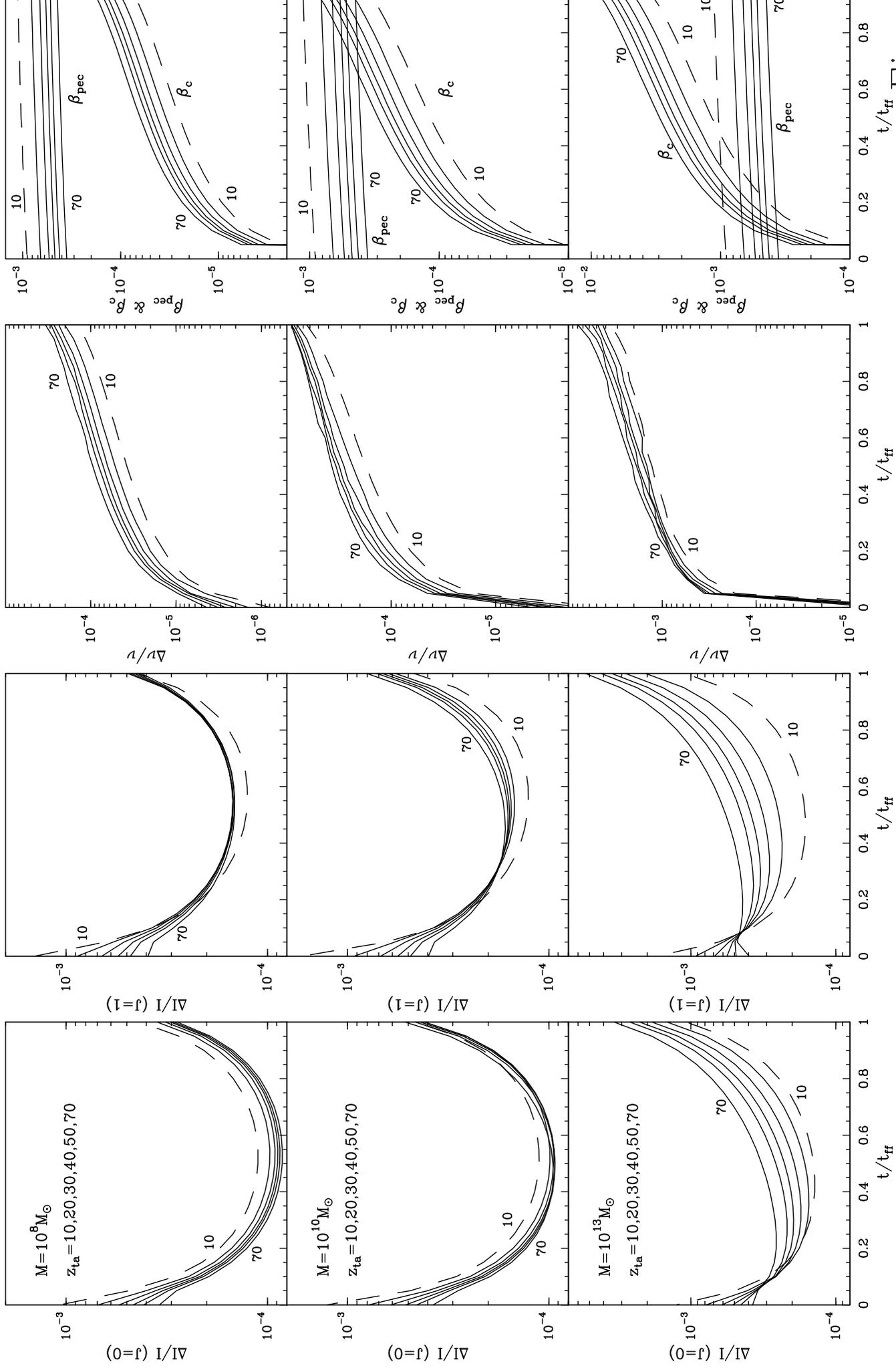

Fig.

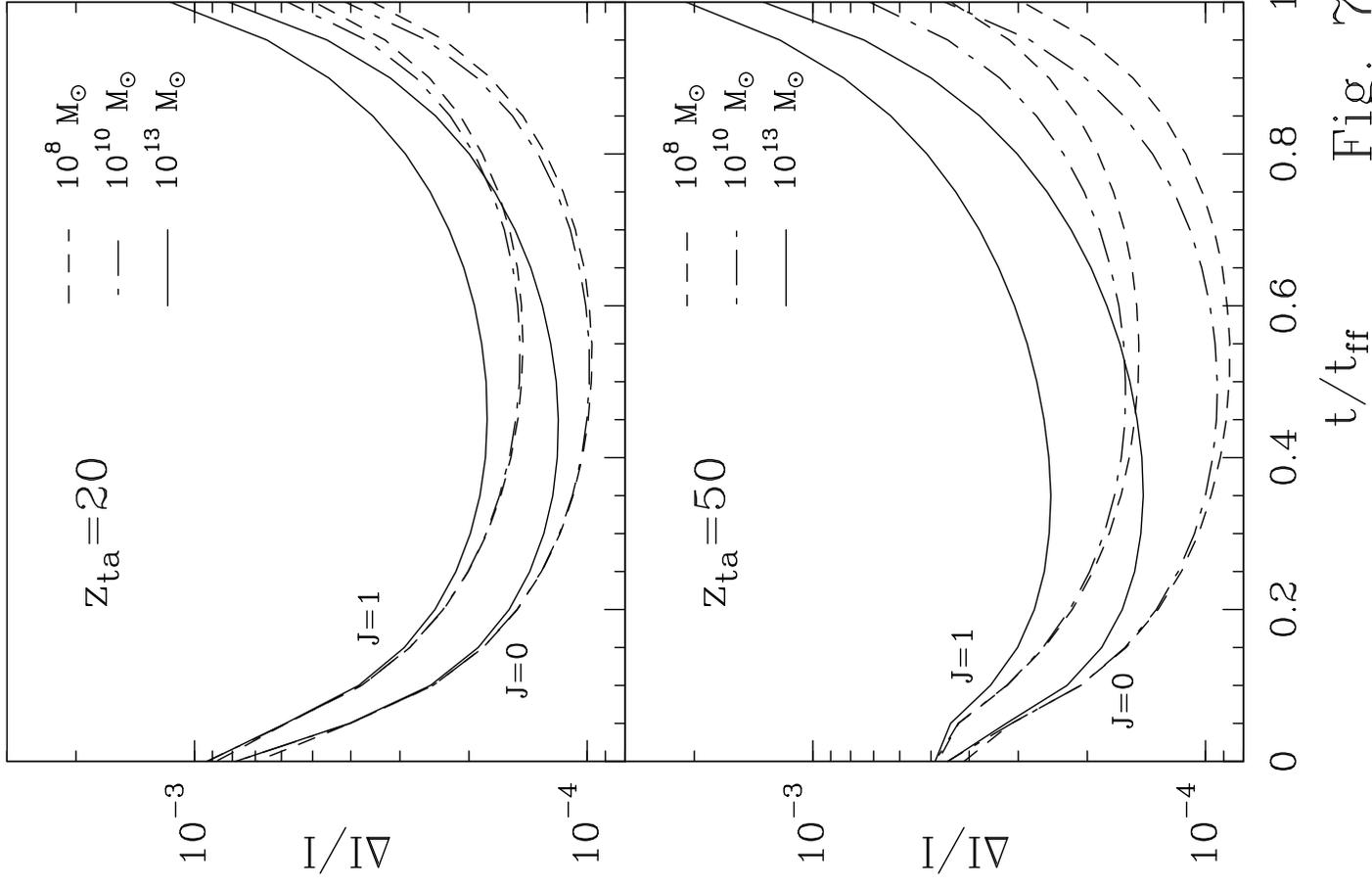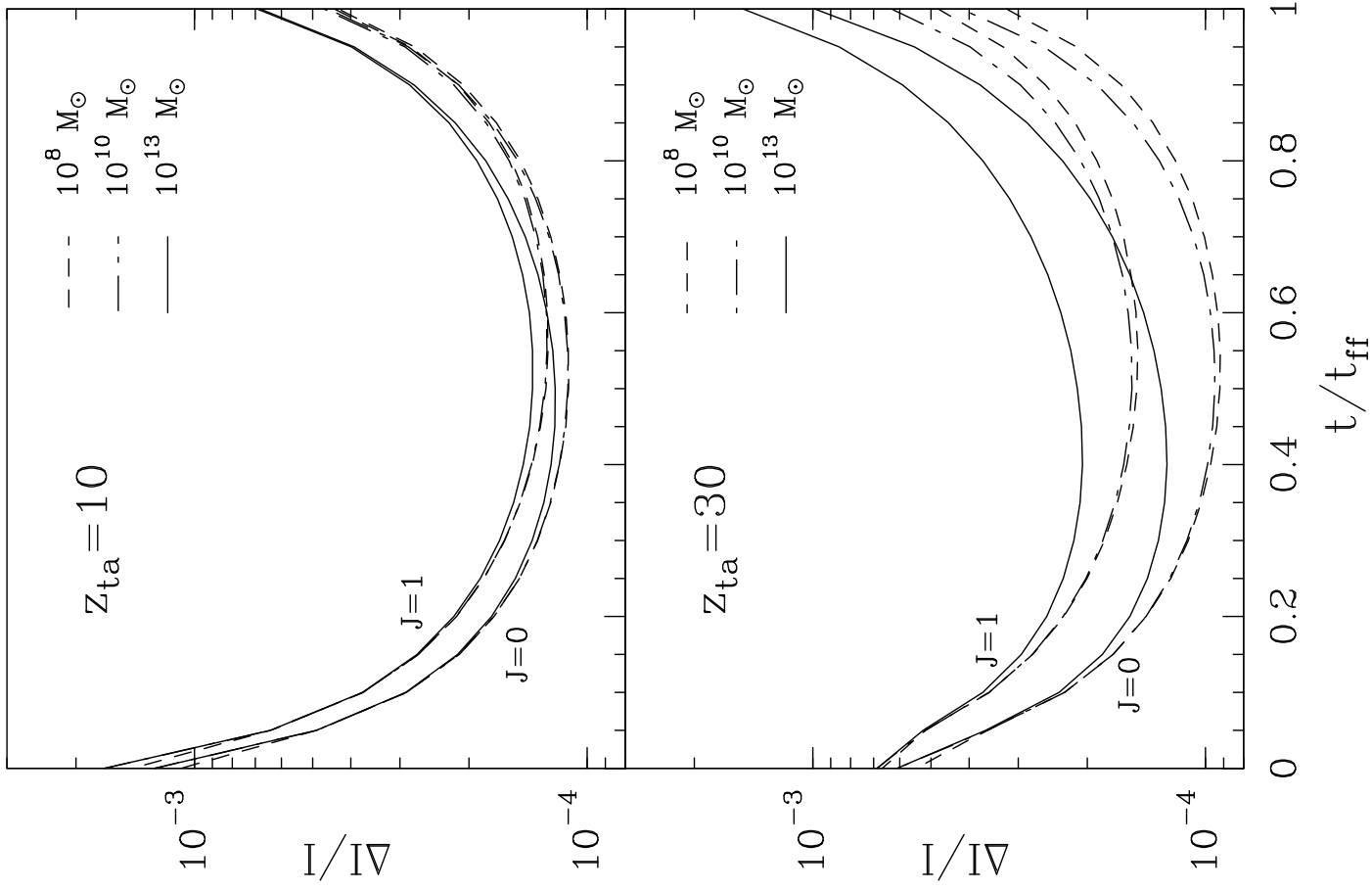

Fig. 7